\documentclass[twocolumn,showpacs,preprintnumbers,superscriptaddress,amsmath,amssymb,prl]{revtex4}%

\usepackage{graphicx} 
\usepackage{dcolumn}  

\graphicspath{{ps}}

\begin{document}


\title { \quad\\[0.5cm]  Measurement of Azimuthal Asymmetries in Inclusive Production\\of Hadron Pairs in $e^+e^-$ Annihilation at Belle}

\affiliation{Budker Institute of Nuclear Physics, Novosibirsk}
\affiliation{Chiba University, Chiba}
\affiliation{Chonnam National University, Kwangju}
\affiliation{University of Cincinnati, Cincinnati, Ohio 45221}
\affiliation{Deutsches Elektronen--Synchrotron, Hamburg}
\affiliation{University of Frankfurt, Frankfurt}
\affiliation{University of Hawaii, Honolulu, Hawaii 96822}
\affiliation{High Energy Accelerator Research Organization (KEK), Tsukuba}
\affiliation{Institute of High Energy Physics, Chinese Academy of Sciences, Beijing}
\affiliation{Institute of High Energy Physics, Vienna}
\affiliation{Institute of High Energy Physics, Protvino}
\affiliation{Institute for Theoretical and Experimental Physics, Moscow}
\affiliation{J. Stefan Institute, Ljubljana}
\affiliation{Kanagawa University, Yokohama}
\affiliation{Korea University, Seoul}
\affiliation{Kyungpook National University, Taegu}
\affiliation{Swiss Federal Institute of Technology of Lausanne, EPFL, Lausanne}
\affiliation{University of Ljubljana, Ljubljana}
\affiliation{University of Maribor, Maribor}
\affiliation{University of Melbourne, Victoria}
\affiliation{Nagoya University, Nagoya}
\affiliation{Nara Women's University, Nara}
\affiliation{National Central University, Chung-li}
\affiliation{National United University, Miao Li}
\affiliation{Department of Physics, National Taiwan University, Taipei}
\affiliation{H. Niewodniczanski Institute of Nuclear Physics, Krakow}
\affiliation{Nippon Dental University, Niigata}
\affiliation{Niigata University, Niigata}
\affiliation{Nova Gorica Polytechnic, Nova Gorica}
\affiliation{Osaka City University, Osaka}
\affiliation{Osaka University, Osaka}
\affiliation{Panjab University, Chandigarh}
\affiliation{Peking University, Beijing}
\affiliation{Princeton University, Princeton, New Jersey 08544}
\affiliation{RIKEN BNL Research Center, Upton, New York 11973}
\affiliation{University of Science and Technology of China, Hefei}
\affiliation{Seoul National University, Seoul}
\affiliation{Shinshu University, Nagano}
\affiliation{Sungkyunkwan University, Suwon}
\affiliation{University of Sydney, Sydney NSW}
\affiliation{Tata Institute of Fundamental Research, Bombay}
\affiliation{Toho University, Funabashi}
\affiliation{Tohoku Gakuin University, Tagajo}
\affiliation{Tohoku University, Sendai}
\affiliation{Department of Physics, University of Tokyo, Tokyo}
\affiliation{Tokyo Institute of Technology, Tokyo}
\affiliation{Tokyo Metropolitan University, Tokyo}
\affiliation{Tokyo University of Agriculture and Technology, Tokyo}
\affiliation{University of Tsukuba, Tsukuba}
\affiliation{Virginia Polytechnic Institute and State University, Blacksburg, Virginia 24061}
\affiliation{Yonsei University, Seoul}
   \author{R.~Seidl\footnote{Permanent address: University of Illinois Urbana
Champaign}}\affiliation{RIKEN BNL Research Center, Upton, New York 11973} 
   \author{K.~Hasuko}\affiliation{RIKEN BNL Research Center, Upton, New York 11973} 
   \author{K.~Abe}\affiliation{Tohoku Gakuin University, Tagajo} 
   \author{I.~Adachi}\affiliation{High Energy Accelerator Research Organization (KEK), Tsukuba} 
   \author{H.~Aihara}\affiliation{Department of Physics, University of Tokyo, Tokyo} 
   \author{D.~Anipko}\affiliation{Budker Institute of Nuclear Physics, Novosibirsk} 
   \author{Y.~Asano}\affiliation{University of Tsukuba, Tsukuba} 
   \author{T.~Aushev}\affiliation{Institute for Theoretical and Experimental Physics, Moscow} 
   \author{A.~M.~Bakich}\affiliation{University of Sydney, Sydney NSW} 
   \author{V.~Balagura}\affiliation{Institute for Theoretical and Experimental Physics, Moscow} 
% ...
\collaboration{The Belle Collaboration}
\noaffiliation
\date{\today}
\begin{abstract}
The Collins effect connects transverse quark spin with a
measurable azimuthal dependence in the yield of hadronic fragments
around the quark's momentum vector. Using two different reconstruction
methods we find evidence of statistically significant azimuthal asymmetries for
charged pion pairs in $e^+e^-$ annihilation at a center-of-mass energy
of 10.52 GeV, which can be attributed to a transverse polarization of
the primordial quarks. 
The measurement was performed using a sample of 79
million hadronic events collected with the Belle detector.
\end{abstract}
\maketitle
The relationship between quark spin and the properties of hadrons
is still poorly understood. 
In the fragmentation of a quark into hadrons, 
Collins~\cite{collins2} has proposed 
a relation between transverse quark spin and the
final state azimuthal distribution of hadrons around the
original quark momentum direction.
Azimuthal asymmetries of hadron yields have recently been reported
in deep inelastic scattering off transversely polarized hydrogen~\cite{hermes} and
deuteron~\cite{compass} targets:
these are believed to be due to the product of
Collins fragmentation functions and
transverse quark spin distribution functions in the nucleon.
However, these so-called transversity distributions,
which contribute to the nucleon transverse spin, 
have not been independently measured. 
Similarly,
no measurement of Collins fragmentation functions has yet been reported.

In this Letter we present a
measurement of azimuthal asymmetries in hadron-hadron correlations for
inclusive charged dihadron production $e^+e^-\rightarrow hhX$, which
we interpret as a direct measure of the Collins effect. Exploratory
studies for this measurement were reported in~\cite{efremov}.
This analysis was performed using a sample of 79 million events
($29\,\mathrm{fb}^{-1}$)
collected at a center-of-mass system (CMS) energy 60 MeV below the
$\Upsilon(4S)$ resonance with the Belle detector~\cite{belledetector}
at the KEKB asymmetric-energy $e^+e^-$ storage rings
\cite{kekb}.  For systematic checks, Monte Carlo (MC) simulated events
generated by the QQ and JETSET \cite{jetset} packages and processed
with a full GEANT-based \cite{geant} simulation of the Belle detector
were used.

The Collins effect occurs in the fragmentation of a quark $q$ with
transverse spin $\mathbf{S_q}$ and 3-momentum $\mathbf{k}$ into an
unpolarized hadron $h$ with transverse momentum $\mathbf{P}_{h\perp}$ with
respect to the original quark direction. The corresponding number density
is defined as~\cite{trento}
\begin{equation}
D_{h\/q^\uparrow}(z,\mathbf{P_{h\perp}})  = D_1^q(z,P_{h\perp}^2) + H_1^{\perp q}(z,P_{h\perp}^2)\frac{(\hat{\mathbf{k}} \times \mathbf{P}_{h\perp})\cdot \mathbf{S}_q}{ zM_h} ,
\label{eq:cdef}
\end{equation}
where $z= \frac{2E_h}{Q}$ is the fractional energy of the hadron
relative to half of the CMS energy $Q$. The first term describes the
spin averaged fragmentation function (FF), and the second, containing
the Collins function $H_1^{\perp q}(z,P_{h\perp}^2)$, depends on the
spin of the quark and produces an asymmetry as it changes sign
if $\mathbf{S_q}$ flips.  The vector product leads to a $\sin(\phi)$
modulation, where $\phi$ is the azimuthal angle between the plane
spanned by the hadron and quark momenta, and the plane spanned by the
quark momentum and the incoming leptons.  Experimentally the quark
direction is approximated by the thrust axis $\hat{\mathbf{n}}$.

\begin{figure}[t]
\begin{center}
\includegraphics[width=7cm]{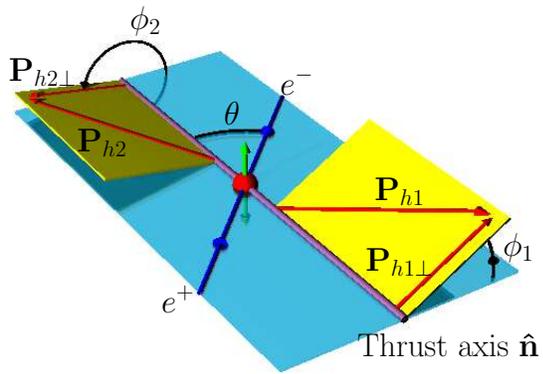}
\caption{Definition of the azimuthal angles of the two hadrons.
In each case, $\phi_i$ is the angle between the plane
spanned by the lepton momenta and the thrust axis $\hat{n}$, and the
plane spanned by $\hat{n}$ and the hadron transverse momentum $P_{h
i \perp}$.}
\label{angle2}
\end{center}
\end{figure}

In hadron production in $e^+e^- \to q\bar{q}$ events, the Collins effect
can be observed when the fragments of the quark and anti-quark are
considered simultaneously. Combining two hadrons from different
hemispheres in jet-like events, with azimuthal angles $\phi_1$ and
$\phi_2$ as defined in Fig.~\ref{angle2} (note that all angular vairablels as well as $\vec{n}$ are defined in the CMS), produces a $\cos( \phi_1 +
\phi_2 )$ modulation of the di-hadron yield. A MC comparison of thrust
axis calculations using reconstructed and generated tracks shows an
average angular deviation between the two of 75 mrad, with a spread with root mean
square of 74 mrad.  This smearing of the reconstructed axis leads to a
reduction in the measured azimuthal asymmetry, as discussed below.

Two experimental methods are used to measure azimuthal
asymmetries. The first method ($M_{12}$) gives rise to the $\cos (
\phi_1 + \phi_2)$ modulation in the di-hadron yields.
The yield is recorded as a function of the hadron angle sum
$\phi_1+\phi_2$, $N_{12}=N_{12}(\phi_1+\phi_2)$, and divided by the
average yield to obtain the normalized rate $R_{12} := N_{12}(\phi_1 +
\phi_2) / \langle N_{12} \rangle$, parametrized by $R_{12} = a_{12}
\cos (\phi_1 + \phi_2) + b_{12}$. Here, $a_{12}$ is a function of the
first moment ($H_1^{\perp q,[1]}$) of the Collins
function~\cite{daniel2}
\begin{equation}
a_{12}(\theta,z_1,z_2)= \frac{\sin^2\theta}{1+\cos^2\theta}\frac{ H_1^{\perp q,[1]}(z_1) \overline{H}{}_1^{\perp q,[1]}(z_2)}{D_1^q(z_1)\overline{D}_1^q(z_2)},
\end{equation}
where $\theta$ is the angle between the incoming lepton axis and the
thrust axis.  An alternative method ($M_{0}$) does not rely on
knowledge of the thrust axis: yields are measured as a function of
$\phi_0$, the angle
between the plane spanned by the momentum vector of the first hadron
and the lepton momenta, and the plane defined by the two hadron
momenta.  The corresponding normalized rate $R_0=N_0(2\phi_0)/\langle
N_0 \rangle$ is a function of $\cos(2\phi_0)$, and (following~\cite{daniel})
can be parametrized as $a_0 \cos(2\phi_0) + b_0$ with
\begin{equation}
a_0(\theta_2,z_1,z_2)= \frac{\sin^2\theta_2}{1+\cos^2\theta_2}\ \frac{f\bigl(H_1^{\perp q}(z_1) \overline{H}_1^{\perp q}(z_2)/M_1M_2\bigr)}{D_1^q(z_1)\overline{D}_1^q(z_2)}~~. 
\end{equation}
$f$ denotes convolution over the transverse hadron momenta. $M_1$
and $M_2$ are the masses of the two hadrons, $z_1$ and $z_2$ are their
fractional energies and $\theta_2$ is the angle between the beam axis
and the second hadron momentum.  The $\sin\theta_2$ dependence
reflects the probability of finding the two initial quarks with
transverse spin.  $\overline{D}^q_1(z)$ and $\overline{H}_1^{\perp q}$
denote fragmentation functions for anti-quarks.


To reduce hard gluon radiation, a two-jet-like topology is
enforced by requiring a thrust value $T>0.8$, 
calculated from all charged and neutral particles with
momentum exceeding $0.1\,\mathrm{GeV}/c$.
The following selection criteria were imposed on the charged pions used in the
analysis methods $M_{12}$ and $M_{0}$: (1) Tracks are
required to originate from the collision vertex,
and to lie in a fiducial region $-0.6 < \cos (\theta_{\rm lab}) < 0.9$,
where $\theta_{\rm lab}$ is the polar angle
in the laboratory frame. 
(2) A likelihood
ratio is used to separate pions from kaons \cite{belledetector}:
$\mathcal{L}(\pi)/[\mathcal{L}(K)+\mathcal{L}(\pi)]>0.7$.  MC
studies show that less than 10\% of pairs have at least one particle
misidentified. (3) We require $z_1, z_2 > 0.2$, to reduce decay
contributions to the pion yields. In addition we require the visible
energy in the detector to exceed $7$ GeV. (4a) The tracks
must lie in opposite jet-hemispheres:
$(\mathbf{P}_{h1}\cdot\hat{\mathbf{n}})
(\mathbf{P}_{h2}\cdot\hat{\mathbf{n}})<0$.  (4b) $Q_T$ is the
transverse momentum of the virtual photon from the $e^+e^-$
annihilation in the rest frame of the hadron pair~\cite{daniel}.
We require $Q_T < 3.5\,\mathrm{GeV}/c$, which
removes contributions from hadrons assigned to the wrong
hemisphere.

\begin{figure}
\includegraphics[width=7cm]{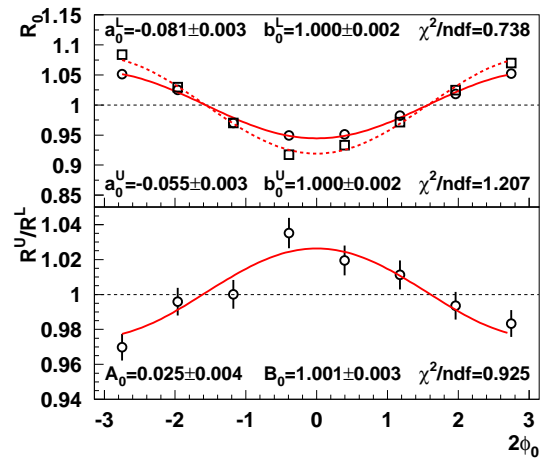}
\caption{\label{fig:raw1}Top: Unlike(U)-sign and like(L)-sign pion pair normalized rate $R_0$ vs.~$2\phi_0$ in the bin $z_1 (z_2) \in
[0.5,0.7]$, $z_2 (z_1) \in [0.3,0.5]$. Bottom: Pion pair
double ratio $R_0^U/R_0^L$ vs.~$2\phi_0$ in the same bin.  The solid and slashed lines show the results of the fit described in the text.}
\end{figure}

The analysis is performed in $(z_1,z_2)$ bins with boundaries at
$z_i=$ 0.2, 0.3, 0.5, 0.7 and 1.0, where complementary off-diagonal bins
$(z_1,z_2)$ and $(z_2,z_1)$ are combined. 
In each $(z_1,z_2)$ bin, 
normalized rates $R_{12}$ and $R_0$ are evaluated in 8 bins of constant
width in the angles $\phi_1 + \phi_2$ and $2\phi_0$ respectively, and
fitted with the functional form introduced above.
Results in the lowest $(z_1,z_2)$ bin are shown in
Fig.~\ref{fig:raw1}. In both methods the constant term ($b_{12}$ or $b_0$)
is found to be consistent with unity for all bins.

In addition to their sensitivity to the Collins effect, $R_{12}$ and
$R_0$ have contributions from instrumental effects and
QCD radiative processes: these are charge independent, and cancel in the double ratio of normalized
rates for unlike-sign(U) over like-sign(L) pion pairs,
\begin{eqnarray}
  \frac{R^U_{\alpha}}{R^L_{\alpha}}
    :=\frac{N^U_{\alpha}(\beta_\alpha)/\langle N^U_{\alpha} \rangle}
           {N^L_{\alpha}(\beta_\alpha)/\langle N^L_{\alpha} \rangle},\nonumber \\
\alpha=0,12, \quad \beta_0 = 2 \phi_0, \quad \beta_{12} = \phi_1 + \phi_2.
\end{eqnarray}
In linear approximation the double ratio is proportional to a combination of the favored and
disfavored fragmentation functions: omitting the transverse
momentum dependence,
\begin{eqnarray}
R_0^U/R_0^L&=&1+ \cos(2\phi_0) \frac{\sin^2\theta}{1+cos^2\theta}\nonumber \\ 
&&\times\Bigg\{\frac{f\left(H_1^{\perp,fav}\overline{H}_1^{\perp,fav} + H_1^{\perp,dis}\overline{H}_1^{\perp,dis}\right)}{ \left(D_1^{fav}\overline{D}_1^{fav} + D_1^{dis}\overline{D}_1^{dis}\right) } \nonumber \\ &&
-\frac{f\left(H_1^{\perp,fav}\overline{H}_1^{\perp,dis}\right)}{ \left(D_1^{fav}\overline{D}_1^{dis}\right)}\Bigg\};
\label{eq:dr}
 \end{eqnarray}
\begin{table*}
\caption{\label{tab:dr} $A_0$ and $A_{12}$ values obtained from fits to pion double ratios as a function of {\it z}. The errors shown are statistical and systematic.}
\begin{ruledtabular}
\begin{tabular}{r l c c } 
\multicolumn{2}{c}{$z_1 \leftrightarrow z_2 $} & $A_0$&$A_{12}$ \\ \hline \hline
$[0.2,0.3]$&$[0.2,0.3]$ & $\hphantom{1}(1.68\pm3.10\pm0.54)\%$&$\hphantom{1}(1.08\pm3.53\pm0.67)\%$ \\ 
$[0.2,0.3]$&$[0.3,0.5]$ & $\hphantom{1}(0.77\pm1.46\pm0.54)\%$&$\hphantom{1}(2.74\pm1.76\pm0.67)\%$ \\ 
$[0.2,0.3]$&$[0.5,0.7]$ & $\hphantom{1}(3.36\pm1.49\pm0.54)\%$&$\hphantom{1}(3.60\pm1.80\pm0.67)\%$ \\ 
$[0.2,0.3]$&$[0.7,1.0]$ & $\hphantom{1}(4.10\pm1.95\pm0.54)\%$&$\hphantom{1}(2.84\pm2.37\pm0.67)\%$ \\ 
$[0.3,0.5]$&$[0.3,0.5]$ & $\hphantom{1}(2.71\pm1.82\pm0.54)\%$&$\hphantom{1}(1.34\pm2.21\pm0.67)\%$ \\ 
$[0.3,0.5]$&$[0.5,0.7]$ & $\hphantom{1}(5.19\pm1.56\pm0.54)\%$&$\hphantom{1}(7.82\pm1.88\pm0.67)\%$ \\ 
$[0.3,0.5]$&$[0.7,1.0]$ & $\hphantom{1}(3.28\pm1.98\pm0.56)\%$&$\hphantom{1}(6.37\pm2.37\pm0.67)\%$ \\ 
$[0.5,0.7]$&$[0.5,0.7]$ & $\hphantom{1}(4.01\pm3.59\pm0.55)\%$&$\hphantom{1}(4.84\pm4.37\pm0.67)\%$ \\ 
$[0.5,0.7]$&$[0.7,1.0]$ & $\hphantom{1}(5.24\pm4.26\pm0.62)\%$&$\hphantom{1}(8.02\pm5.19\pm0.70)\%$ \\ 
$[0.7,1.0]$&$[0.7,1.0]$ & $(12.78\pm6.98\pm0.62)\%$&$(21.84\pm8.34\pm1.59)\%$ \\
\end{tabular}
\end{ruledtabular}
\end{table*}

an analogous expression can be given for $R_{12}^U/R_{12}^L$.  Favored
fragmentation processes (\emph{e.g.}\ $D_1^{fav}$) are transitions in which a
valence quark of the hadron is of the same flavor as the
initial quark, for example $u, \bar{d} \rightarrow \pi^+$;  
the corresponding unfavored process is $u, \bar{d}
\rightarrow \pi^-$.  Following our analysis of the normalized rates we
parametrize the double ratios as $R^U_\alpha/R^L_\alpha= A_{\alpha}
\cos (\beta_{\alpha}) + B_{\alpha}$ with $\alpha=0,12$,
$\beta_0=2\phi_0$ and $\beta_{12}=\phi_1+\phi_2$; the parameters
$A_{\alpha}$, $B_{\alpha}$ are determined from fits in
each $(z_1,z_2)$-bin.

\begin{figure}
\includegraphics[width=7cm]{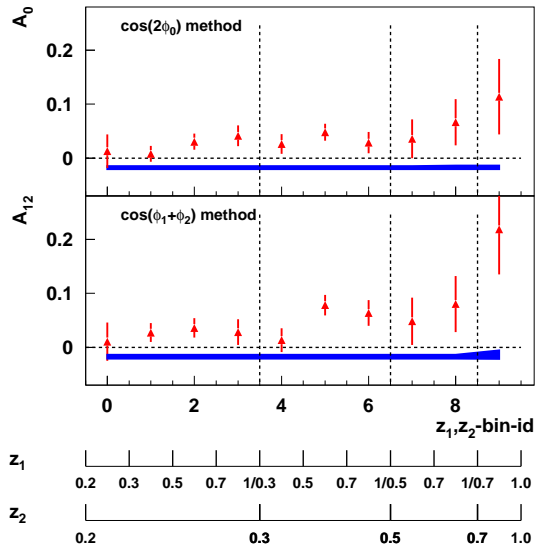}
\caption{\label{fig:pipi}Values of $A_0$ and $A_{12}$
as functions of {\it z},
corrected for the contribution of charm events. The lower scales show
the boundaries of the bins in $z_1$ and $z_2$; see the text. The
shaded band shows the size of the systematic errors.}
\end{figure}
An example is shown in Fig.~\ref{fig:raw1}. The parameters $A_0$ and $A_{12}$, related to the Collins fragmentation functions 
(see Eq.~\ref{eq:dr}),
are shown in Fig.~\ref{fig:pipi} and listed in Table \ref{tab:dr}.  Significant non-zero
values are observed, especially at high $z$. The
observed increase with $z$ is predicted in several models
\cite{alessandro0,alessandro2,leonard}. The weighted averages
over all $(z_1,z_2)$-bins, corrected for the charm
contribution (see below), are found to be $A_0=(3.06\pm0.65\pm0.55)$\%
and $A_{12}=(4.26\pm0.78\pm0.68)$\%\footnote{The raw, charm uncorrected asymmetries are $a_0^U=(-8.90\pm0.07)\%$, $a_0^L=(-10.33\pm0.08)\%$, $a_{12}^U=(5.37\pm0.07)\%$ and $a_{12}^L=(3.14\pm0.08)\%$.}.

By measuring asymmetries using double ratios, acceptance
effects cancel, while the contribution of gluon radiation cancels only
to first order. The size of any additional contribution was estimated
by subtracting like-sign from unlike-sign pair rates, in which
case gluon effects should cancel exactly and only experimental effects
should remain.  
The differences, $0.04$\% on average for method $M_0$ and
$0.03$\% for $M_{12}$, are assigned as systematic errors.
Double ratios were formed using MC events, which
do not include the Collins effect but take into account gluon
radiation and detector effects: the parameters $A_0$ and $A_{12}$ were found to
be consistent with zero. The statistical errors of these fits,
0.45\% and 0.56\% respectively, are included in
the systematic uncertainty.
Double ratios of positively-charged over negatively-charged pairs
were found to be consistent with unity, and limit the
possibility of charge dependent detector effects; 
again, the precisions of the fits ($0.26$\% and $0.21$\%) are included as
systematic errors.
An additional test was performed by taking pion pairs from jet
hemispheres in different events: no asymmetry was found.

In addition to fitting the double ratios with a $\cos\phi$ modulation,
higher harmonics ($\sin 2\phi,\cos 4\phi$) were introduced.
The change in the results,
$0.03$\% on average for $A_0$ and $0.01$\% for $A_{12}$,
was included in the systematic uncertainty. In
method $M_{12}$, the single hadron yield around the thrust axis was also
studied. Although the Collins effect leads to a sinusoidal modulation, this
should average to zero in the absence of a specified quark spin. The
$\sin\phi_{1,2}$ modulation was found to be consistent with
zero.

\begin{figure}
\includegraphics[width=8cm]{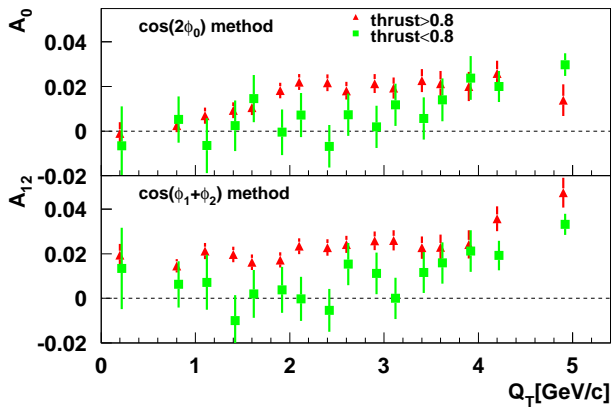}
\caption{\label{fig:thrust} Values of
$A_0$ and $A_{12}$ as a function of $Q_T$ (not
corrected for charm background) for the data samples with
$T>0.8$ (triangles) and $T<0.8$ (squares).}
\end{figure}

As a consistency check, we also measure double ratios from an event sample with a reversed thrust selection,
$T<0.8$: asymmetries in this sample are reduced for $Q_T < 3.5$ GeV/$c$ (see Fig.~\ref{fig:thrust}). For the low-thrust sample there is no clear
two-jet topology, and, thus, the Collins effect should be suppressed
while any radiative or acceptance effects remain present. A fit to a
constant for $Q_T<3.5$ GeV/$c$ finds $(0.4\pm0.3)$\% for both $A_0$
and $A_{12}$.
The asymmetries for $Q_T>3.5$ GeV/$c$ and $T<0.8$  were found to be due
to events with low visible energy as well as background from
interactions in the detector material.

The uncertainty due to particle misidentification was estimated by
applying tighter selection criteria. The difference in the double
ratio with respect to the default selection is included in the
systematic error, $0.12$\% for method $M_0$
and $0.28$\% for $M_{12}$.

To test the reconstruction of azimuthal asymmetries,
Monte Carlo samples were reweighted in $\cos(2\phi_0)$ and
$\cos(\phi_1 + \phi_2)$, producing generated moments of $5$\% and $10$\% for
unlike-sign pairs, and $0$\% and $-5$\% for like-sign pairs.  The
reconstructed azimuthal asymmetries were consistent with generated
values for method $M_0$, but lower by $17.6\pm 1.1$\% on average for
method $M_{12}$, which depends on the thrust axis of the reconstructed
event. (Note that the thrust axis represents the direction of the
outgoing quarks only on average.) This dilution was corrected by
rescaling $A_{12}$ with a factor $1.210 \pm 0.014$.

Both methods depend on the assumption that the electron and positron
beams are unpolarized. This was tested by studying the angular
distribution of $e^+e^-\rightarrow \mu^+ \mu^-$ events: no significant
azimuthal asymmetries are observed and thus no systematic error assigned to it.  We correct for the contribution
from charm decays using measured asymmetries in events where a $D^*$
meson has been reconstructed, and the $e^+e^-\rightarrow c\bar{c}$
event fraction determined from MC (23\%). The corresponding uncertainty is
        included in the statistical errors in Table~\ref{tab:dr}. The
        fraction of selected events due to the $e^+e^-\rightarrow \tau^+\tau^-$ process is small (1.7\%), and the asymmetries obtained in a
tau enhanced data sample have low statistical significance: $A_0=(0.17\pm0.30)\%$ and $A_{12} =(0.74\pm0.30)\%$. This contribution is 
added to the systematic error. All systematic uncertainties were added in quadrature.
Correlations among individual angular and $(z_1,z_2)$-bins were
tested using a large number of MC samples, by comparing uncertainties returned by the fits with the expected values. We
find the statistical error to be underestimated by 14\%;
the final uncertainty is increased by the corresponding factor.

In summary we have performed a measurement of the azimuthal asymmetry
in the inclusive production of pion pairs as a function of the
fractional energy {\it z} of pion pairs. In the double ratio of
asymmetries from unlike-sign and like-sign pairs, possible
contributions of gluon radiation and detector effects cancel and the
observed asymmetry can be attributed to the Collins effect.  This asymmetry in $e^+e^-$ annihilation is the first direct evidence for this effect.

\begin{acknowledgments}
The authors would like to thank D.~Boer for fruitful discussions on the theoretical aspects of the measurement. 
We thank the KEKB group for the excellent operation of the
accelerator, the KEK cryogenics group for the efficient
operation of the solenoid, and the KEK computer group and
the NII for valuable computing and Super-SINET network
support.  We acknowledge support from MEXT and JSPS (Japan);
ARC and DEST (Australia); NSFC (contract No.~10175071,
China); DST (India); the BK21 program of MOEHRD and the CHEP
SRC program of KOSEF (Korea); KBN (contract No.~2P03B 01324,
Poland); MIST (Russia); MHEST (Slovenia);  SNSF (Switzerland); NSC and MOE
(Taiwan); and DOE and NSF (USA).
\end{acknowledgments}


\end{document}